\documentstyle[aps]{revtex}

\begin{document}
\draft
\preprint{HEP/123-qed}
\title{Effect of a Domain Wall on the Conductance Quantization 
        in a Ferromagnetic Nanowire}
\author{Katsuyuki Nakanishi and Yoshiko Oi Nakamura\cite{Nakamura}}
\address{Department of Applied Physics, Science University of Tokyo, 
        Kagurazaka, Shinjuku-ku, Tokyo 162, Japan}
\maketitle
\begin{abstract}
The effect of the domain wall (DW) on the conductance 
in a ballistic ferromagnetic nanowire (FMNW) 
is revisited by exploiting a specific perturbation theory
which is effective for a thin DW; the thinness is often the case
in currently interested conductance measurements on FMNWs.
Including the Hund coupling 
between carrier spins and local spins in a DW, 
the conductance of a FMNW in the presence of a very thin DW 
is calculated within the Landauer-B\"{u}ttiker formalism.
It is revealed that the conductance plateaus are modified
significantly,
and the switching of the quantization unit from $e^2/h$ 
to ``about $2e^2/h$'' is produced in a FMNW
by the introduction of a thin DW.
This accounts well for recent observations in a FMNW. 
\end{abstract}

Owing to technical development in nanofabrication 
and spin-controlled measurement, 
much interest has attracted recently to the transport phenomena 
in FMNWs and ferromagnetic nanocontacts (FMNCs).
Whether the effect of the magnetic DWs on the resistivity 
in a ferromagnetic wire is positive or negative is a problem 
which has been argued for a long time from both experimental 
and theoretical sides~\cite{PR165,PRL77,PRB51,ITM34,PRL80,PSS61,PRB60,PRL78,PRB53,PRL79,JMS23,PRL82,PRL83,cm100,PRB59},
and still remains a matter of controversy.
This controversy stems from the fact 
that the effect is affected by and entangled with various factors 
in actual quantum ferromagnetic wires; 
the presence of impurities, the band structures 
and the size of the contacts 
as well as the experimental geometries and conditions.
Recently, relating with device technologies,
intensive research efforts have been focused on the conductance 
in FMNWs and FMNCs~\cite{APL75,PRB55}.
In most of these FMNWs, the length of the contacts are shorter 
than the electronic mean free path 
so that the transport can be regarded as ballistic 
and the DW is restricted in a very narrow region.
Among such works, a recent one of the conductance measurement
on a high quality Ni nanocontact,
which is stretched into a nanowire~\cite{APL75},
reported that a distinct staircase behavior is observed 
just before the wire breaks.
Further the step height of the staircase changes 
from $2e^2/h$ like to $e^2/h$ by an application 
of parallel magnetic fields to the wire axis beyond the saturation 
during this elongation process.
It may be understood that this switching would occur 
when a DW present in the case of lower fields is eliminated 
by the application of the saturation field 
and the magnetization is ferromagnetically saturated (FMS) 
along the wire axis.
But it is not so obvious why the quantization unit of the conductance 
in the presence of a DW becomes ``about $2e^2/h$'';
$2e^2/h$ is the quantization unit 
in the degenerate diamagnetic nanowires.

In this paper, being inspired by the these observations, we make 
a theoretical study on the conductance of a FMNW with a thin DW
in the ballistic regime.
In the zero field case of the measurement mentioned above,
the conductance looks to follow so perfect $2e^2/h$ step staircase 
at the last stage before the wire breaks that the nanowire may 
be supposed to be in the ballistic regime. 
Therefore, our study in the ballistic regime may fit suitably 
with the experiment, and is expected to give an explanation 
to the origin of the observed change in the quantization unit 
of the conductance,
if the origin ever traces back to the presence of a DW.

The common ingredient of the ballistic electron transport 
in quantum wires is Landauer-B\"{u}ttiker formula~\cite{PM21,PRB31}, 
which gives the conductance $G$ as

\begin{eqnarray}
        G = \frac{e^2}{h} \sum_{n} \sum_{\sigma}
                t_{\sigma} (E_{\|}(n)), \label{G}
\end{eqnarray}

\noindent
where $E_{\|}(n)$ is the energy of the longitudinal motion 
of conduction electrons in the $n$th channel,
and $t_{\sigma}(E_{\|}(n))$ is the corresponding 
transmission probability of the incident electrons 
with spin $\sigma (= \uparrow,\downarrow)$. 
In early studies~\cite{PSS61},
the resistance arising from the electronic scattering by a DW
of ordinary thickness in a pure magnetic wire was calculated
as to be exponentially small although it is positive,
and the contribution was shown later 
in an adiabatic approximation~\cite{PRB60} 
to decrease quadratically in the inverse DW width.
In those works, however, the metallic wire is thick enough 
for the longitudinal energy $E_{\|}(n)$ to be taken 
as $E_{F}$ (Fermi energy) in most of channels,
and the thickness of the DW is large enough 
so that electrons track adiabatically the exchange field in the DW
and therefore the backward scattering by the DW is negligibly small.
In nanoscale wires, on the other hand, the confinement 
of electrons in the transverse direction forces 
to open only a restricted number of channels,
in most of which $E_{\|}(n)$ could become small.
If we assume a perfect confinement, for simplicity, 
$E_{\|}(n)$ is given by 

\begin{eqnarray}
        E_{\|}(n) = E_{\rm tot} - E_{n} 
        \equiv E_{F} - \frac{\hbar^2}{2m} 
        \biggl(\frac{\pi n}{W} \biggr)^2, \label{energy}
\end{eqnarray} 

\noindent
where $W$ is the transverse dimension of the wire, 
$m$ is the mass of electrons, and  
the total energy $E_{\rm tot}$ is equated to $E_{F}$.

We can show, in the following calculation,
that $t_{\sigma}(E_{\|})$ is notably spin-dependent 
as well as deviates from unity significantly
when $E_{\|}$ goes to the low energy comparable 
with the exchange energy $V_0$ 
between electronic spins and local spins in a DW.
Now, in a wire of nanoscale width, the number 
of opening channels decreases and $E_{\|}(n)$ in most of channels
goes into the low energy region, where the effect 
of the electronic scattering by the DW on the conductance becomes relevant. 
Then the effect is expected
to give a significant modification to the conductance plateaus.
This is our scenario to explain the change in the staircase behavior
of the conductance appearing in the presence of a DW.

We begin with the following effective Hamiltonian for electrons 
of one-dimensional conduction along $z$-axis across a $180^{\circ}$
DW of width $2\lambda$;

\begin{eqnarray}
        H = - \frac{\hbar^2}{2m} \frac{d^2}{dz^2}
                -V_{0} \tanh \biggl( \frac{z}{\lambda} \biggr)
                \ \sigma_{z}
                -V_{0} {\rm sech} \biggl( \frac{z}{\lambda} \biggr)
                \ \sigma_{x},
        \label{Hamiltonian}
\end{eqnarray}

\noindent
where the DW is centered at $z=0$ and $\sigma_{i}$ are the Pauli matrices.
The second and the third terms are the exchange potential 
and the spin-flip potential felt by electrons respectively. 
This Hamiltonian is the same one which was introduced first 
by Cabrera and Falicov considering the effective coupling 
between the electronic spins with the local magnetization 
in a DW~\cite{PSS61}.
This is also derived in a recent work~\cite{JMS23} 
as an effective Hamiltonian for electrons 
interacting with quantum spins in the DW by the Hund coupling. 
This Hamiltonian mixes the spin channels, so that, 
to calculate the transmission probabilities $t_{\sigma}(E_{\|})$ 
for an electron 
to go from $z=-\infty$ to $z=\infty$ with energy $E_{\|}$ across the DW,
we are forced to solve a one-dimensional two component 
Shr\"{o}dinger equation

\begin{eqnarray}
         H \Psi_{\sigma}(z) = E_{\|} \Psi_{\sigma}(z), \label{equation}
\end{eqnarray}

\noindent
where the index $\sigma(= \uparrow,\downarrow)$ denotes the spin state 
of the incident electron 
and $\Psi_{\sigma}(z)$ is two component column vector such 
as $\Psi_{\sigma}(z)=( \psi_{\uparrow \sigma}(z),
\psi_{\downarrow \sigma}(z) )$ for each $\sigma = \uparrow,\downarrow$.
This coupled Schr\"{o}dinger equations are difficult 
to be solved analytically and have been never solved successfully.
Recently it is reported that the equation can be solved analytically
for the case of the sinusoidal form of potentials~\cite{cm406}.
However, the assumption of such a potential form produces artifacts
such as oscillatory behaviors in $t_{\sigma}(E_{\|})$.
To avoid this, we like to solve Eq.(\ref{equation}) keeping 
with the potential forms in Eq.(\ref{Hamiltonian}) but perturbationally. 
There are two ways in perturbational approach 
which are complementary each other:
One is valid for a thick DW where the unperturbed state is 
that of a electron tracking adiabatically the local field in the DW,
so that the perturbation represents the mistracking.
This is the usual way employed in literatures~\cite{PRL78,PRL79,PRL83}.
The other, on the other hand, is valid for a thin DW.
There, a state of complete mistracking appears as an unperturbed state.
We proceed in the latter way. 
This specific perturbational method is made possible 
by having solved exact Green's function 
$G_{\sigma}^{0}(z,z^{\prime};E_{\|})$ 
corresponding to the Hamiltonian $H_{0}$, which has  
a step-like potential 
$v_{0}(z) \equiv -V_0 \{ \theta(z)-\theta(-z) \} \sigma_{z}$;

\begin{eqnarray}
         H_{0} = -\frac{\hbar^2}{2m} \frac{d^2}{dz^2}
                 +v_0(z)\sigma_{z}
\end{eqnarray}

\noindent
Then we have $H=H_0+H_1$ with $H_1$ given by 

\begin{eqnarray}
         H_{1} &=& \Biggr\{-V_0 \tanh \biggl( \frac{z}{\lambda} \biggr)
                 -v_0(z) \Biggl\} \sigma_z
                 -V_0{\rm sech} \biggl( \frac{z}{\lambda} \biggr)
                 \ \sigma_x \nonumber \\
             &\equiv & v_1(z)\sigma_z + v_2(z)\sigma_x. \label{perturbed}
\end{eqnarray}

\noindent
Both of $v_1(z)$ and $v_2(z)$ have finite values 
only in a region $|z| {< \atop \sim} \lambda $, 
so that $H_1$ can be dealt with as a perturbation.

The Schr\"{o}dinger equations (\ref{equation})
are put into the Lippmann-Schwinger form;

\begin{eqnarray}
         \Psi_{\sigma}(z) = \Phi_{\sigma}(z) 
         +\int_{-\infty}^{\infty} dz^{\prime}
                 G^0(z,z^{\prime};E_{\|}) H_1(z^{\prime}) 
                 \Psi_{\sigma}(z^{\prime}), \label{lseq}
\end{eqnarray}

\noindent
where $\Phi_{\sigma}(z)$ are scattering solutions 
of the unperturbed Schr\"{o}dinger equation 
with the step function potential
$(H_{0} \Phi_{\sigma}(z) = E_{\|} \Phi_{\sigma}(z))$.
The unperturbed Green's function 
$G^{0}(z,z^{\prime};E_{\|})$ is 
a $2 \times 2$ diagonal matrix whose diagonal elements are 
$G_{\uparrow}^{0}(z,z^{\prime};E_{\|})$ 
and $G_{\downarrow}^{0}(z,z^{\prime};E_{\|})$; 
their explicit forms are given in Appendix.
We solve Eq.(\ref{lseq}) up to the 2nd order in $H_1$.
The perturbation expansion of Eq.(\ref{lseq}) 
is written in terms of dimensionless quantities
$\tilde{z}=\frac{z}{\lambda}$, $\tilde{H}_1$ and $\tilde{G}^0$ defined
by $H_1(z)=V_0\tilde{H}_1(\frac{z}{\lambda})$ and
$G^0(z_2,z_1;E_{\|})=\frac{\gamma}{2iV_0\lambda}
\tilde{G}^0(\frac{z_2}{\lambda},\frac{z_1}{\lambda};E_{\|})$
with $\gamma=(2m\lambda^2V_0/\hbar^2)^{1/2}$;

\begin{eqnarray}
  &\Psi_{\sigma}&(\lambda \tilde{z}) = 
  \Phi_{\sigma}(\lambda \tilde{z})
  + \frac{\gamma}{2i} \int_{-\infty}^{\infty} \!\!\!\!\!\! d\tilde{z}_1
    \tilde{G}^0(\tilde{z},\tilde{z}_1;E_{\|})\tilde{H}_1(\tilde{z}_1)
    \Phi_{\sigma}(\lambda\tilde{z}_1) \nonumber \\
  &+& \!\! \biggl( \frac{\gamma}{2i} \biggr)^2
    \!\! \int_{-\infty}^{\infty} \!\!\!\!\!\! d\tilde{z}_1
    \!\!\! \int_{-\infty}^{\infty} \!\!\!\!\!\!  d\tilde{z}_2
    \tilde{G^0}(\tilde{z},\tilde{z}_1;E_{\|})\tilde{H}_1(\tilde{z}_1) 
    \nonumber \\
  & &~~~~~~~~~~~~~~~~~~~\times 
    \tilde{G^0}(\tilde{z}_1,\tilde{z}_2;E_{\|})\tilde{H}_1(\tilde{z}_2)
    \Phi_{\sigma}(\lambda\tilde{z}_2)
  + \cdots .
\end{eqnarray}

Since $\int_{-\infty}^{\infty} \!\!\!  d\tilde{z}^{\prime}
\tilde{G^0}(\tilde{z},\tilde{z}^{\prime};E_{\|})
\tilde{H_1}(\tilde{z}^{\prime})\approx O(1)$ 
as $\tilde{H}_1(\tilde{z}) \approx O(1)$ 
for $|\tilde{z}| {< \atop \sim} 1$,
the small parameter of the expansion is $\gamma$.
The transmission and the reflection coefficient 
$S_{\sigma \sigma^{\prime}}$ and $R_{\sigma \sigma^{\prime}}$
are obtained from the asymptotic forms;

\begin{eqnarray}
    \Psi_{\uparrow}(z) \equiv \left(
            \begin{array}{c}
                    \psi_{\uparrow \uparrow}(z)\\
                    \psi_{\downarrow \uparrow}(z)
            \end{array} \right)
          & \raisebox{-1.9mm}{\overrightarrow{\; z \rightarrow -\infty \;}} & 
            \left(
            \begin{array}{l}
                    e^{ik_2z}+R_{\uparrow \uparrow} e^{-ik_2z} \\
                    R_{\downarrow \uparrow} e^{-ik_1z}
            \end{array} \right) \nonumber \\
          & \raisebox{-1.9mm}{\overrightarrow{\; z \rightarrow +\infty \;}} &
            \left(
            \begin{array}{l}
                    S_{\uparrow \uparrow} e^{ik_1z} \\
                    S_{\downarrow \uparrow} e^{ik_2z}
            \end{array} \right) \\
            \label{upwf}
    \Psi_{\downarrow}(z) \equiv \left(
            \begin{array}{c}
                    \psi_{\uparrow \downarrow}(z)\\
                    \psi_{\downarrow \downarrow}(z)
            \end{array} \right)
          & \raisebox{-1.9mm}{\overrightarrow{\; z \rightarrow -\infty\;}} &
            \left(
            \begin{array}{l}
                    R_{\uparrow \downarrow} e^{-ik_2z} \\
                    e^{ik_1z}+R_{\downarrow \downarrow} e^{-ik_1z}
            \end{array} \right) \nonumber \\
          & \raisebox{-1.9mm}{\overrightarrow{\; z \rightarrow +\infty\;}} &
            \left(
            \begin{array}{l}
                    S_{\uparrow \downarrow} e^{ik_1z} \\
                    S_{\downarrow \downarrow} e^{ik_2z}
            \end{array} \right) 
            \label{downwf}
\end{eqnarray}

\noindent
where $\hbar k_{1} = \sqrt{2m(E_{\|}+V_0)}$ and 
$\hbar k_{2} = \sqrt{2m(E_{\|}-V_0)}$.

The transmission probabilities $t_{\sigma}(E_{\|})$ 
for the incident electron 
with spin $\sigma(=\uparrow,\downarrow)$ to transmit 
to any final spin state are calculated by relations 
$t_{\uparrow}(E_{\|})= \frac{k_1}{k_2} 
         | S_{\uparrow \uparrow}(E_{\|}) |^2
         + | S_{\downarrow \uparrow}(E_{\|}) |^2$ and 
$t_{\downarrow}(E_{\|})= \frac{k_2}{k_1} 
         | S_{\downarrow \downarrow}(E_{\|}) |^2
         + | S_{\uparrow \downarrow}(E_{\|}) |^2$.
The results are shown in FIG.\ \ref{fig1} 
for appropriate values of $\lambda$.
These figures show that (1) $t_{\sigma}(E_{\|})$ deviates 
from unity significantly only for energies comparable 
with $V_0$,
and (2) the spin-flip transmission probability rises linearly 
in $E_{\|}/V_0$ at the threshold $E_{\|}/V_0=-1$,
and the gradient is about $4\pi^2\gamma^2$ for small $\gamma$.
This means that, the thinner the DW is, 
the harder the spin-flip transmission occurs.
The last point indicates that electron spins become hard
to track local spins in a thin DW adiabatically.

Now we can easily find 
how the conductance plateaus of a FMNW
are modified by the presence of a thin DW. 
The $W$-dependence of the conductance is derived from the formula;

\begin{eqnarray}
         G=\frac{e^2}{h} \Biggl[ \sum_{n=1}^{N_{\uparrow}} 
                 t_{\uparrow}(E_{\|}(n)) + 
                 \sum_{n=1}^{N_{\downarrow}} 
                 t_{\downarrow}(E_{\|}(n)) \Biggr] 
           \equiv G_{\uparrow}+G_{\downarrow}.
\end{eqnarray}

\noindent
Numbers $N_{\uparrow}$ and $N_{\downarrow}$ are defined 
by $N_{\uparrow}= \bigl[ W \{ 2m(E_F - V_0)/ \pi^2 \hbar^2 \}^{1/2} \bigr]$
and $N_{\downarrow}= \bigl[ W \{ 2m(E_F + V_0)/ \pi^2 \hbar^2 \}^{1/2} \bigr]$
respectively, where the square bracket denotes the Gauss symbol.
The curve of $G$ versus $W$ in the presence of a single DW 
(thick solid line) is shown in FIG.\ \ref{fig2}(a) 
together with those of $G_{\uparrow}$ (thin solid line)
and $G_{\downarrow}$ (thin dashed line).
In FIG.\ \ref{fig2}(b), the corresponding curves for the case of FMS 
are drawn for comparison. 
In the case of FMS, the exchange energy felt 
by $\uparrow$-spin and $\downarrow$-spin electrons 
differs by $2V_0$. Due to this difference, the threshold value 
of $W_{\sigma n}$, at which the $n$th channel 
opens, becomes different between the $\uparrow$-spin and
the $\downarrow$-spin channel;  
$W_{\downarrow n} = n \{ \pi^2 \hbar^2 / 2m(E_{F}+V_0) \}^{1/2}$ 
and 
$W_{\uparrow n} = n \{ \pi^2 \hbar^2 / 2m(E_{F}-V_0) \}^{1/2}$.
The opening of the channel begins with the first $\downarrow$-spin one 
and is followed by the first $\uparrow$-spin one, and so on.
This leads to the $e^2/h$ conductance staircase with clear plateaus
in FIG.\ \ref{fig2}(b). This $e^2/h$ staircase behavior of the conductance 
in the case of FMS are observed 
in conductance measurements on Ni nanowires~\cite{APL75,PRB55}. 
Now we discuss the result in the presence of a DW shown 
in FIG.\ \ref{fig2}(a).
There, the conductance curve looks like a staircase 
with step height of ``about $2e^2/h$''.
While the step height is well quantized at least up to the 3rd step, 
steps are gradually inclined as the steps increases
although a remnant of $2e^2/h$ like quantization is still seen. 
Our curve in FIG.\ \ref{fig2}(a) resembles one 
observed recently~\cite{APL75} in this characteristic appearance.
Therefore, we can say that the switching of the quantization unit 
from $e^2/h$ to ``about $2e^2/h$'' can be produced 
by the introduction of a thin DW into a FMNW.

Although it is the case, 
we should make a remark about the origin of the conductance 
quantization in the unit of ``about $2e^2/h$''. 
It is sometimes said that the presence of a DW
makes both spin channels stand on the equal footing, and as a result 
the spin degeneracy is recovered as it stands in the diamagnetic nanowire.
Precisely speaking, it is not exactly the case. 
By looking at curves
$G_{\uparrow}$ and $G_{\downarrow}$ in FIG.\ \ref{fig2}(a),
we see that they are quite different from each other. 
By comparing $G_{\uparrow}$ and $G_{\downarrow}$ in FIG.\ \ref{fig2}(a)
with the correspondings in FIG.\ \ref{fig2}(b), 
we notice the following: (1) $G_{\uparrow}$ jumps by an amount 
$e^2/h$ at the same value of $W_{\uparrow n}$ 
as $G_{\uparrow}$ does in the FMS case,
although the corners of the staircase are rounded by the scattering effect 
by the DW. 
(2) The $n$th channel of $G_{\downarrow}$ opens also at the same value 
of $W_{\downarrow n}$ as in the FMS case.
There, however $G_{\downarrow}$ does not jump up as in the FMS case,
since only the slowly increasing spin-flip transmission 
contributes first to $G_{\downarrow}$ (see FIG.\ \ref{fig1}(b)).
When $W$ reaches the threshold value of the spin-conserving transmission
in the same channel, $G_{\downarrow}$ starts to increase steeply 
by this contribution.
This threshold coincides with $W_{\uparrow n}$
of $G_{\uparrow}$, so that the total conductance $G$ looks to jump up 
by about $2e^2/h$ at $W_{\uparrow n}$.
Speaking in another way, since the Hamiltonian~(\ref{Hamiltonian}) 
without the third spin-flip term is invariant 
under the reflection $z \rightarrow -z$ 
and the $\pi$ rotation about the $x$-axis in the spin space,
we can easily find that the spin-conserving transmission probability
for $\uparrow$-spins and $\downarrow$-spins become exactly the same.
Further, both of $W_{\sigma n}$ becomes equal 
to $n \{ \pi^2 \hbar^2 / 2m(E_{F}-V_0) \}^{1/2}$,
so that the total conductance realizes a staircase 
of exact $2e^2/h$ steps.
In fact, our perturbational calculation 
for this case (FIG.\ \ref{fig2}(c)) shows 
a exactly quantized behavior with fine plateaus.
In such case, therefore, we can say that the perfect quantization 
in the unit of $2e^2/h$ precisely comes from the spin degeneracy.
>From these considerations, we understand 
that the ``about $2e^2/h$'' staircase behavior in the presence 
of a thin DW
does not indicate precisely the recovery of the spin degeneracy.
We can even claim that the deviation 
of the curve in FIG.\ \ref{fig2}(a) 
from the perfect $2e^2/h$ staircase shows clearly a distinct effect 
of the spin-flip forward scattering by a DW.

In summary, we studied theoretically the effect 
of a thin DW on the conductance quantization in a FMNW 
in the ballistic regime. 
The calculations are made 
by exploiting a specific perturbational technique 
which is valid for thin DW case.
We point out that notable effect can be observed owing to the nanoscale
width of the wire.
We also find that the conductance quantization in the unit 
of ``about $2e^2/h$'' is realized by the introduction 
of a thin $180^{\circ}$ DW
while the quantization unit is $e^2/h$ 
in the saturated ferromagnetic nanowire as ordinarily expected.
This explains well the switching in the quantization unit
from ``about $2e^2/h$'' to $e^2/h$ observed recently 
in measurements on Ni nanowires~\cite{APL75}.
It is emphasized that, in the deviation of the ``$2e^2/h$ like'' 
conductance staircase from the perfect $2e^2/h$ one in lack 
of the spin-flip scattering,
we can find a novel effect of the spin-flip scattering by the DW,
from which we can deduce useful informations for the ferromagnetic material.
In this sense, the conductance measurement in a FMNW
may provide a powerful probe for the magnetotransport 
in ferromagnetic materials. 

\

The authors thank Masanori Yamanaka and Atsuo Satou 
for helpful comments.
We are grateful to Teruo Ono 
for informing us their observations in an early stage prior 
to the publication, and for useful discussions.

\appendix
\section{}
The unperturbed Green's function $G_{\sigma}^0(z_2,z_1;E_{\|})$
can be obtained by using some mathematics as;

\begin{eqnarray}
  &G&_{\downarrow}^0(z_2,z_1;E_{\|}) = \frac{m}{i\hbar^2} \nonumber \\
  & &\times
  \left\{
    \begin{array}{ll}
      \frac{1}{k_2}e^{i k_2 z_2}(e^{-i k_2 z_1}-Re^{i k_2 z_1}) &
      ;~~z_2>z_1>0 \\
      \frac{1}{k_2}(e^{-i k_2 z_2}-Re^{i k_2 z_2})e^{i k_2 z_1} &
      ;~~z_1>z_2>0 \\
      \frac{2}{k_1+k_2}e^{i k_2 z_2}e^{-i k_1 z_1} &
      ;~~z_2>0>z_1 \\
      \frac{2}{k_1+k_2}e^{-i k_1 z_2}e^{i k_2 z_1} &
      ;~~z_1>0>z_2 \\
      \frac{1}{k_1}(e^{i k_1 z_2}+Re^{-i k_1 z_2})e^{-i k_1 z_1} &
      ;~~0>z_2>z_1 \\
      \frac{1}{k_1}e^{-i k_1 z_2}(e^{i k_1 z_1}+Re^{-i k_1 z_1}) &
      ;~~0>z_1>z_2,
    \end{array} \right. \\
  &G&_{\uparrow}^0(z_2,z_1;E_{\|})=G_{\downarrow}^0(-z_1,-z_2;E_{\|}),
\end{eqnarray}

\noindent
where $R=(k_1-k_2)/(k_1+k_2)$.


\figure{
 \caption{The transmission probabilities
         (a) $t_{\uparrow}$ and (b) $t_{\downarrow}$ 
         as functions of $E_{\|}/V_0$
         with $V_0=0.001$ eV. \label{fig1}}}

\figure{
 \caption{The conductance as a function of $W$ in cases; 
         (a) involving a single domain wall,
         (b) of FMS,
         and (c) without the spin-flip scattering.
         There we take $V_0=0.001$ eV, $E_{F}=10V_0$ and 
         $\lambda=12.0$ \AA. \label{fig2}}}


\begin{references}

 \bibitem[*]{Nakamura}
 To whom the correspondence is to be addressed. 
 E-mail address: nakamura@grad.ap.kagu.sut.ac.jp

 \bibitem{PR165}
 G.R. Taylor, A. Isin and R.V. Coleman, Phys. Rev. {\bf 165}, 621 (1968).

 \bibitem{PRL77}
 J.F. Gregg, W. Allen, K. Ounadjela, M. Viret, M. Hehn, S.M. Thompson 
 and J.M.D. Coey, Phys. Rev. Lett. {\bf 77}, 1580 (1996).

 \bibitem{PRB51}
 K. Hong and N. Giordano, Phys. Rev. B{\bf 51}, 9855 (1995).

 \bibitem{ITM34}
 Y. Otani, S.G. Kim and K. Fukamichi, IEEE Trans. Magn. {\bf 34}, 1096 (1998).

 \bibitem{PRL80}
 U. Ruediger, J. Yu, S. Zhang, A.D. Kent and S.S.P. Parkin, 
 Phys. Rev. Lett. {\bf 80}, 5639 (1998).

 \bibitem{PSS61}
 G.G. Cabrera and L.M. Falicov, 
 Phys. Status Solidi (b) {\bf 61}, 539 (1974), ibid {\bf 62}, 217 (1974).

 \bibitem{PRB60}
 A. Brataas, G. Tatara and G.E.W. Bauer,
 Phys. Rev. B{\bf 60}, 3406 (1999).

 \bibitem{PRL78}
 G. Tatara and H. Fukuyama, Phys. Rev. Lett. {\bf 78}, 3773 (1997).

 \bibitem{PRB53}
 M. Viret, D. Vignoles, D. Cole, J.M.D. Coey, W. Allen, 
 D.S. Daniel and J.F. Gregg,
 Phys. Rev. B{\bf 53}, 8464 (1996).

 \bibitem{PRL79}
 P.M. Levy and S. Zhang, Phys. Rev. Lett. {\bf 79}, 5110 (1997).

 \bibitem{JMS23}
 M. Yamanaka and T. Koma, J. Magn. Soc. Japan {\bf 23}, 141 (1999).

 \bibitem{PRL82}
 N. Garc\'{\i}a, M. Mu\~{n}oz and Y.W. Zhao,
 Phys. Rev. Lett. {\bf 82}, 2923 (1999).

 \bibitem{PRL83}
 G. Tatara, G.W. Zhao, M. Mu\~{n}oz and N. Garc\'{\i}a,
 Phys. Rev. Lett. {\bf 83}, 2030 (1999).

 \bibitem{cm100}
 R.P. van Gorkom, A. Brataas and G.E.W. Bauer,
 cond-mat/9910100.

 \bibitem{PRB59}
 J.B.A.N. van Hoof, K.M. Schep, A. Brataas, G.E.W. Bauer and
 P.J. Kelly,
 Phys. Rev. B{\bf 59}, 138 (1999).
 
 \bibitem{APL75}
 T. Ono, Y. Ooka, H. Miyajima and Y. Otani, 
 Appl. Phys. Lett. {\bf 75}, 1622 (1999).

 \bibitem{PRB55}
 J.L. Costa-Kr\"{a}mer, Phys. Rev. B{\bf 55}, R4875 (1997).

 \bibitem{PM21}
 R. Landauer,
 Philos. Mag. {\bf 21}, 863 (1970).

 \bibitem{PRB31}
 M. B\"{u}ttiker, Y. Imry, R. Landauer, S. Pinhas,
 Phys. Rev. B{\bf 31}, 6207 (1985).

 \bibitem{cm406}
 H. Imamura, N. Kobayashi, S. Takahashi and S. Maekawa,
 cond-mat/9906406.

\end{references}
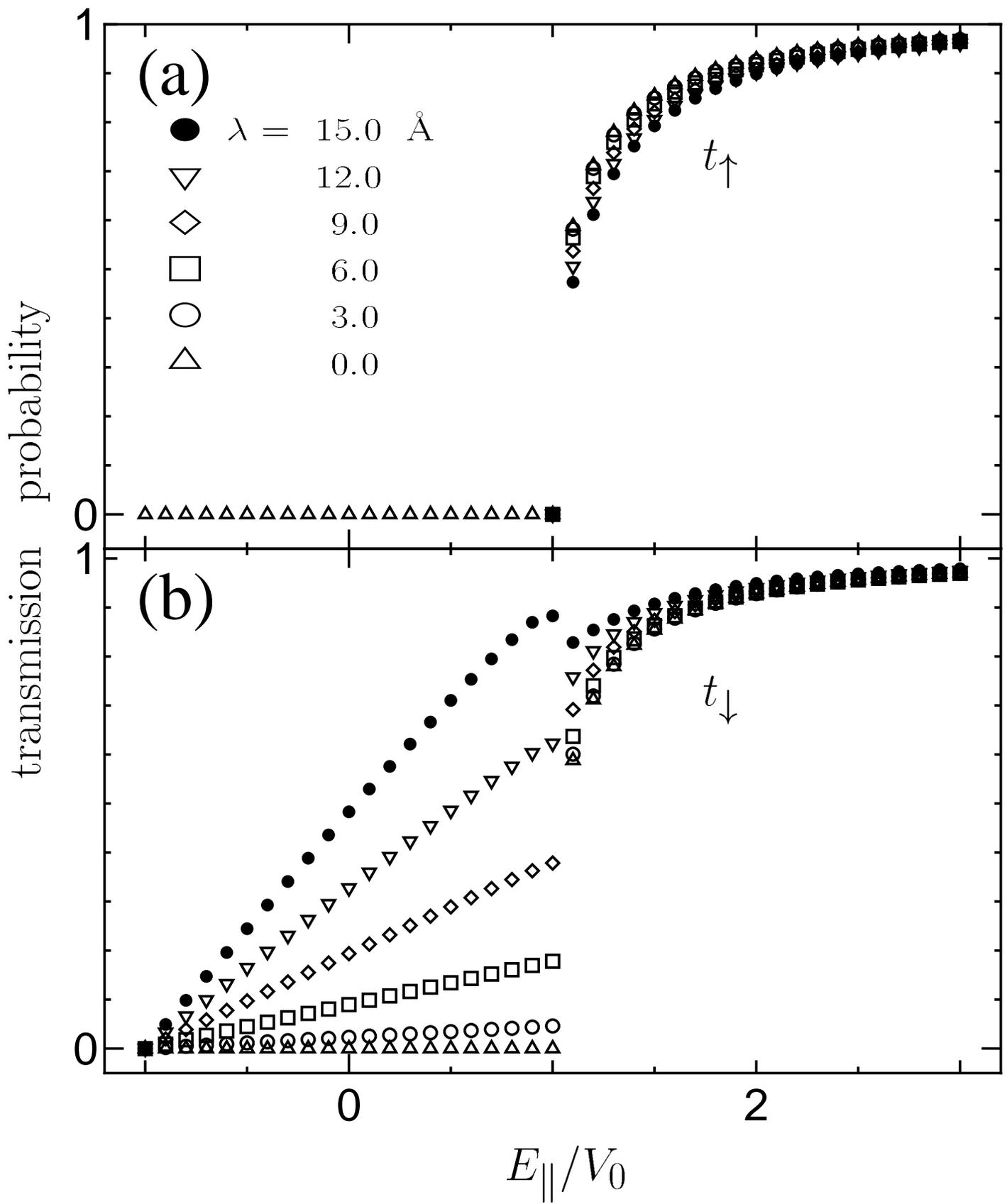
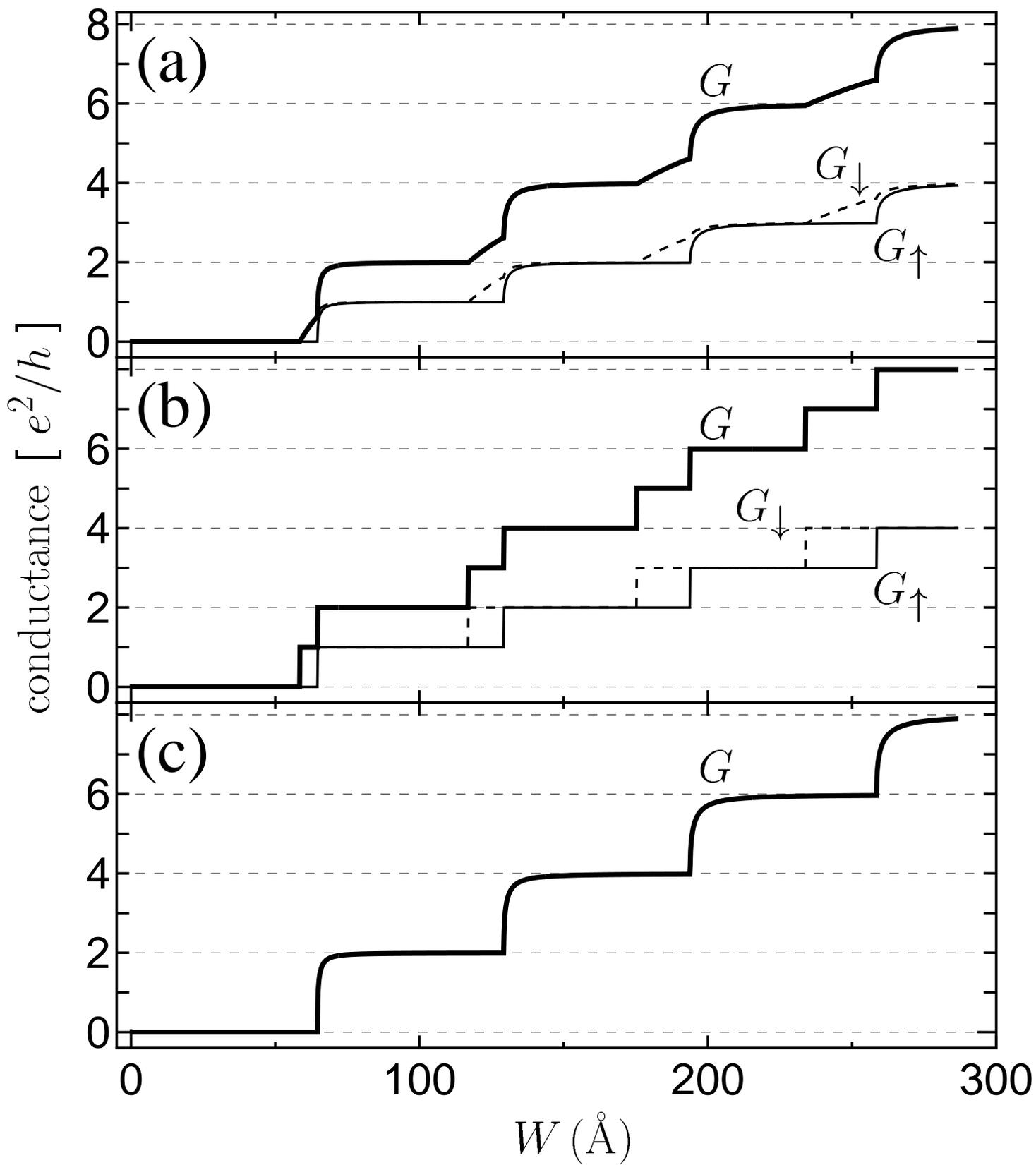
\end{document}